\newcommand{\CLASSINPUTtoptextmargin}{19mm}%
\newcommand{\CLASSINPUTbottomtextmargin}{43mm}%
\newcommand{\CLASSINPUTinnersidemargin}{12.9mm}%
\newcommand{\CLASSINPUToutersidemargin}{12.9mm}%
\documentclass[conference,10pt,a4paper]{IEEEtran}
\usepackage{amsmath}
\usepackage{times}
\usepackage{graphicx}
\usepackage{multirow}
\usepackage[none]{hyphenat}
\usepackage{float}
\usepackage{subfig}

\usepackage{url}

\usepackage{stfloats}
\usepackage[abbreviations]{glossaries-extra}

\glssetcategoryattribute{acronym}{indexonlyfirst}{true}

\newabbreviation{cnn}{CNN}{Convolutional Neural Network}
\newabbreviation{rnn}{RNN}{Recurrent Neural Network}
\newabbreviation{mse}{MSE}{Mean Squared Error}
\newabbreviation{snr}{SNR}{Signal-to-Noise Ratio}

\makeatletter

\def\@maketitle{\newpage
\bgroup\par\addvspace{0.5\baselineskip}\centering%
\ifCLASSOPTIONtechnote
   {\bfseries\large\@IEEEcompsoconly{\sffamily}\@title\par}\vskip 1.3em{\lineskip .5em\@IEEEcompsoconly{\sffamily}\@author
   \@IEEEspecialpapernotice\par{\@IEEEcompsoconly{\vskip 1.5em\relax
   \@IEEEtitleabstractindextextbox{\@IEEEtitleabstractindextext}\par
   \hfill\@IEEEcompsocdiamondline\hfill\hbox{}\par}}}\relax
\else
   \vskip0.2em{\EuMWtitlesize\ifCLASSOPTIONtransmag\bfseries\LARGE\fi\@IEEEcompsoconly{\sffamily}\@IEEEcompsocconfonly{\normalfont\normalsize\vskip 2\@IEEEnormalsizeunitybaselineskip
   \bfseries\Large}\@title\par}\vskip1.0em\par
   \ifCLASSOPTIONconference%
      {\@IEEEspecialpapernotice\mbox{}\vskip\@IEEEauthorblockconfadjspace%
       \mbox{}\hfill\begin{@IEEEauthorhalign}\@author\end{@IEEEauthorhalign}\hfill\mbox{}\par}\relax
   \else
      \ifCLASSOPTIONpeerreviewca
         {\@IEEEcompsoconly{\sffamily}\@IEEEspecialpapernotice\mbox{}\vskip\@IEEEauthorblockconfadjspace%
          \mbox{}\hfill\begin{@IEEEauthorhalign}\@author\end{@IEEEauthorhalign}\hfill\mbox{}\par
          {\@IEEEcompsoconly{\vskip 1.5em\relax
           \@IEEEtitleabstractindextextbox{\@IEEEtitleabstractindextext}\par\hfill
           \@IEEEcompsocdiamondline\hfill\hbox{}\par}}}\relax
      \else
         \ifCLASSOPTIONtransmag
           {\@IEEEspecialpapernotice\mbox{}\vskip\@IEEEauthorblockconfadjspace%
            \mbox{}\hfill\begin{@IEEEauthorhalign}\@author\end{@IEEEauthorhalign}\hfill\mbox{}\par
           {\vspace{0.5\baselineskip}\relax\@IEEEtitleabstractindextextbox{\@IEEEtitleabstractindextext}\vspace{-1\baselineskip}\par}}\relax
         \else
           {\lineskip.5em\@IEEEcompsoconly{\sffamily}\sublargesize\@author\@IEEEspecialpapernotice\par
           {\@IEEEcompsoconly{\vskip 1.5em\relax
            \@IEEEtitleabstractindextextbox{\@IEEEtitleabstractindextext}\par\hfill
            \@IEEEcompsocdiamondline\hfill\hbox{}\par}}}\relax
         \fi
      \fi
   \fi
\fi\par\addvspace{0.0\baselineskip}\egroup}

\def\EuMWtitlesize{\@setfontsize{\EuMWtitlesize}{24}{24pt}}
\def\EuMWauthorsize{\@setfontsize{\EuMWauthorsize}{11}{11pt}}
\def\EuMWaffilsize{\@setfontsize{\EuMWaffilsize}{10}{10pt}}
\def\EuMWcaptionsize{\@setfontsize{\EuMWcaptionsize}{9}{10pt}}
\def\EuMWbibsize{\@setfontsize{\EuMWbibsize}{8}{10pt}}

\def\@IEEEauthorblockNstyle{\EuMWauthorsize\@IEEEcompsocnotconfonly{\sffamily}\@IEEEcompsocconfonly{\large}}
\def\@IEEEauthorblockAstyle{\EuMWaffilsize\@IEEEcompsocnotconfonly{\sffamily}\@IEEEcompsocconfonly{\itshape}\@IEEEcompsocconfonly{\large}}
\def\@IEEEauthordefaulttextstyle{\EuMWauthorsize\@IEEEcompsocnotconfonly{\sffamily}\sublargesize}

\def\thebibliography#1{\section*{\refname}%
    \addcontentsline{toc}{section}{\refname}%
    \EuMWbibsize\@IEEEcompsocconfonly{\small}\vskip 0.3\baselineskip plus 0.1\baselineskip minus 0.1\baselineskip
    \list{\@biblabel{\@arabic\c@enumiv}}%
    {\settowidth\labelwidth{\@biblabel{#1}}%
    \leftmargin\labelwidth
    \advance\leftmargin\labelsep\relax
    \itemsep \IEEEbibitemsep\relax
    \usecounter{enumiv}%
    \let\p@enumiv\@empty
    \renewcommand\theenumiv{\@arabic\c@enumiv}}%
    \let\@IEEElatexbibitem\bibitem%
    \def\bibitem{\@IEEEbibitemprefix\@IEEElatexbibitem}%
\def\newblock{\hskip .11em plus .33em minus .07em}%
\ifCLASSOPTIONtechnote\sloppy\clubpenalty4000\widowpenalty4000\interlinepenalty100%
\else\sloppy\clubpenalty4000\widowpenalty4000\interlinepenalty500\fi%
    \sfcode`\.=1000\relax}

\long\def\@makecaption#1#2{%
\ifx\@captype\@IEEEtablestring%
\par\@IEEEtabletopskipstrut
\else
\@IEEEfigurecaptionsepspace
\fi
\setbox\@tempboxa\hbox{\normalfont\footnotesize {#1.}\nobreakspace\nobreakspace #2}%
\ifdim \wd\@tempboxa >\hsize%
\setbox\@tempboxa\hbox{\normalfont\footnotesize {#1.}\nobreakspace\nobreakspace}%
\parbox[t]{\hsize}{\normalfont\footnotesize\noindent\unhbox\@tempboxa#2}%
\else
\ifCLASSOPTIONconference \hbox to\hsize{\normalfont\footnotesize\hfil\box\@tempboxa\hfil}%
\else \hbox to\hsize{\normalfont\footnotesize\box\@tempboxa\hfil}%
\fi\fi
\ifx\@captype\@IEEEtablestring%
\@IEEEtablecaptionsepspace
\else
\fi}

\newlength\tablecaptiontotableskip
\newlength\figuretocaptionskip
\def\@IEEEfigurecaptionsepspace{\vskip\figuretocaptionskip\relax}%
\def\@IEEEtablecaptionsepspace{\vskip\tablecaptiontotableskip\relax}%

\def\abstract{\normalfont%
\@IEEEabskeysecsize\bfseries\textit{\abstractname}\,\bfseries\textit{---}\,%
\@IEEEgobbleleadPARNLSP}%

\def\IEEEkeywords{\normalfont%
\@IEEEabskeysecsize\bfseries\textit{\IEEEkeywordsname}\,\bfseries\textit{---}\,%
\@IEEEgobbleleadPARNLSP}%
\def\endIEEEkeywords{\relax\vspace{0.67ex}%
\par\if@twocolumn\else\endquotation\fi%
\normalsize\normalfont}%

\def\@IEEEauthorblockNtopspace{0ex}
\def\@IEEEauthorblockAtopspace{1mm}
\def\tablename{Table}
\def\thetable{\arabic{table}}
\def\IEEEkeywordsname{Keywords}
\def\subsubsection{\@startsection{subsubsection}{3}{\z@}{1.5ex plus 1.5ex minus 0.5ex}%
{0.7ex plus .5ex minus 0ex}{\normalfont\normalsize\itshape}}%
\newlength{\CPheadmatchindent}%
\def\@seccntformat#1{\hbox to\CPheadmatchindent{\csname the#1dis\endcsname}\hskip 0.1em \relax}
\IEEEilabelindentA \parindent
\IEEEilabelindent \IEEEilabelindentA
\IEEEelabelindent \parindent
\IEEEdlabelindent \parindent
\IEEElabelindent \parindent
\makeatother

\begin{document}
\raggedbottom
%
%
%
\title{A Novel Micro-Doppler~Coherence~Loss for~Deep~Learning Radar~Applications}

%
%
\author{\IEEEauthorblockN{
Mikolaj~Czerkawski,
Christos~Ilioudis,
Carmine~Clemente,
Craig~Michie,
Ivan~Andonovic,
Christos~Tachtatzis
}
\IEEEauthorblockA{%
Department of Electronic and Electrical Engineering, University of Strathclyde, UK\\
}
}
%
\maketitle
%
%
\begin{abstract}
Deep learning techniques are subject to increasing adoption for a wide range of micro-Doppler applications, where predictions need to be made based on time-frequency signal representations. Most, if not all, of the reported applications focus on translating an existing deep learning framework to this new domain with no adjustment made to the objective function. This practice results in a missed opportunity to encourage the model to prioritize features that are particularly relevant for micro-Doppler applications. Thus the paper introduces a micro-Doppler coherence loss, minimized when the normalized power of micro-Doppler oscillatory components between input and output is matched. The experiments conducted on real data show that the application of the introduced loss results in models more resilient to noise.
\end{abstract}
%
%

\section{Introduction}
    Advancements in deep learning have motivated diverse research in their application to radar signal processing since many of the relevant tasks can be described as instances of classification or domain translation. The paper reports on the enhanced performance of deep learning models processing Doppler-time radar signals, achieved by changing the content of the objective function.

    Although a number of deep learning approaches have been applied to radar signal processing, research to adjust the objective function within this domain has been limited. In \cite{Kim2016}, one of the early applications of a \gls{cnn} for the classification of radar signals is reported and although a performance improvement is evident owing to deep learning techniques, the type of loss used is not revealed. The unsupervised approach utilising a stacked auto-encoder architecture presented in \cite{Jokanovic2016} utilised a cost function consisting of a reconstruction term, based on the \gls{mse}, a weight regularisation term, and a divergence term between a sparsity parameter and the average output of the hidden neurons. A \gls{cnn} model for human activity classification is introduced in \cite{Lang2017}, but the exact characteristics of the objective function are not disclosed. A transfer learning approach using a novel DivNet architecture for human motion classification is proposed in \cite{Seyfioglu2018}, without modification of the objective function content. An unsupervised approach to learning relevant features from radar micro-Doppler spectrograms in \cite{Seyfioglu2018a} applies a standard auto-encoder objective function. A significant body of research related to the classification of radar signals has adopted the same methodologies \cite{Shrestha2019, Erol2019, Erol2019a, Alnujaim2020, Wang2019, Huizing2019, Gurbuz2019, Li2019}.

     Here, a micro-Doppler coherence loss (an additional term within the objective function) is introduced, applicable in a wide range of frameworks operating on radar Doppler signals. The micro-Doppler coherence loss improves results in an unsupervised learning scheme applied for a classification task by promoting aligned periodic characteristics of the reconstructed signal in individual velocity bands to that of the ground truth signal. Results indicate that deep neural networks performing tasks related to micro-Doppler analysis can achieve superior immunity to injected noise when trained using this loss.
     
\section{Problem Formulation}
    
    Two representative application contexts are selected to position the scope of the reported results. First, a network executing domain translation where a given time-frequency map is transformed to a different domain - de-noising or interference removal fall into this category; second is classification, one of the common uses of deep learning for radar applications.
    
    In general, each application requires the network to learn relevant features from the input time-frequency map. In the case of domain translation, the features establish a representation used to decode an appropriate output, while in the case of classification, these features form the input to a classification output module (for instance, a relatively shallow stack of fully connected layers). The feature set learned by the encoding module significantly impacts the performance of the network in both cases. Consequently, the challenge of learning the relevant features constitutes the problem.
    
    Furthermore, reliance on the standard losses widely adopted outside of the radar context can advertise features that are not relevant in the goal of interpreting Doppler signals. The most commonly applied objective function $J_\theta$ for fully convolutional networks with parameters $\theta$, is the \gls{mse} reconstruction loss between the target $y$ and network output $\hat{y}$ (both of size $M\times N$):
    \begin{equation}
        J_\theta(y, \hat{y}) = \mathcal{L}_{\textrm{MSE}}(y, \hat{y}) = \frac{1}{N}\frac{1}{M}
        \sum_{f_D=1}^{N}\sum_{t=1}^{M}(y_{t,f_D}-\hat{y}_{t,f_D})^2
        \label{eq:pixel_loss}
    \end{equation}
    This loss does not prioritise any set of high-level features over others in the sense that the value of each pixel has an equal and direct influence on the error. The choice of prioritised features is challenging, but if the assumption that the relevant information is not uniformly spread over all pixels holds true, then an improvement in performance can be expected through adjustment of the loss terms.
    
    \begin{figure}
        \centering
        \includegraphics[width=\columnwidth]{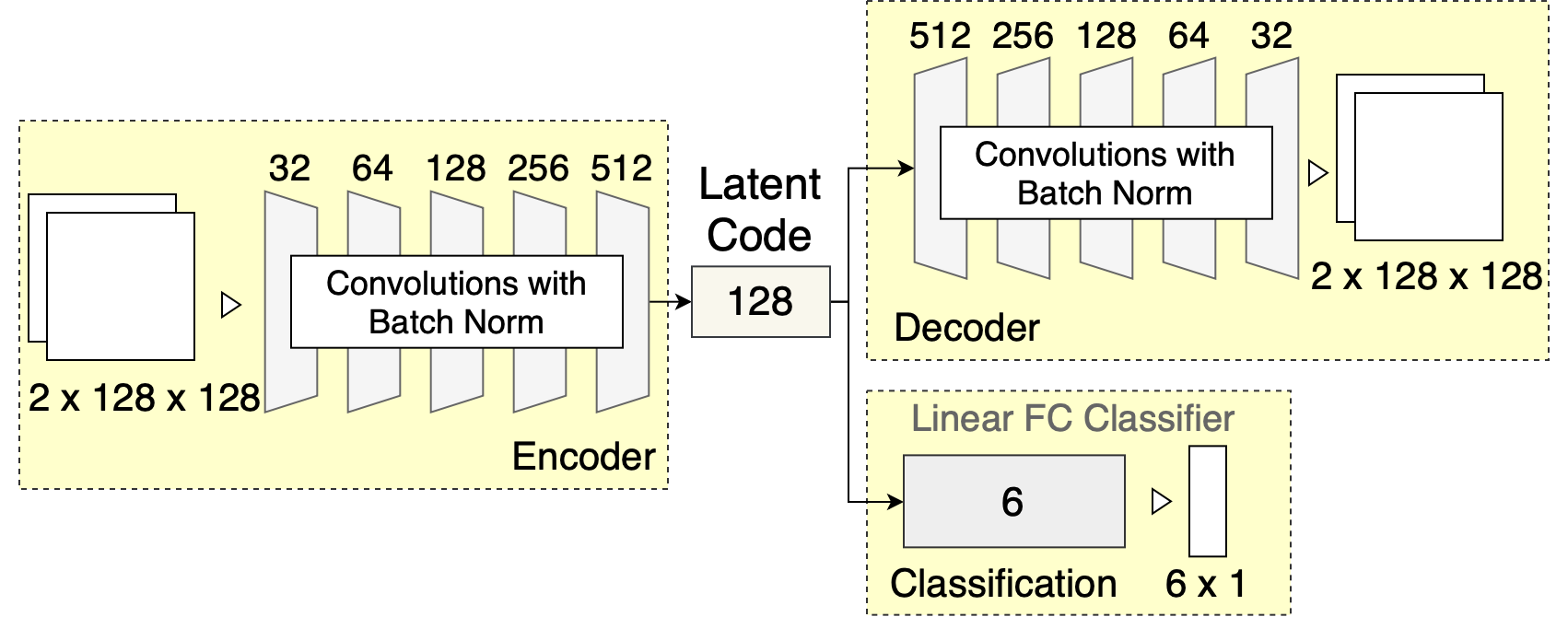}
        \caption{Diagram of the used hybrid model.}
        \label{fig:model_diagram}
    \end{figure}
    
\section{Proposed Solution}
    The modulations of Doppler components contain crucial information in micro-Doppler applications. Consequently, uniform contribution of each time-frequency bin to the total error may not be appropriate. A new loss term of micro-Doppler coherence loss $\mathcal{L}_{\mu\textrm{D}}$ with a weight $\beta$ is proposed to promote the relevant micro-Doppler features. The introduced loss term is designed to promote spectral similarity within each Doppler band. This is achieved by formulating a metric that will be minimized when the normalized spectral distribution of each Doppler band is the same for both compared signals.
    
    \begin{equation}
        J_\theta(y, \hat{y}) = \mathcal{L}_{\textrm{MSE}}(y, \hat{y}) + \beta\cdot \mathcal{L}_{\mu\textrm{ D}}(y, \hat{y})
        \label{eq:j_updated}
    \end{equation}
    
    For $\mathcal{L}_{\mu\textrm{ D}}$, both a 2D time-frequency ground truth matrix $y[t,f_D]$ and a corresponding network output $\hat{y}[t,f_D]$ are subject to discrete Fourier transform $\mathcal{F}_t$ applied in the temporal dimension $t$, transforming it to cadence frequency $t \to f_c$. This yields two tensors representing Doppler-cadence maps (still of size $M\times N$). The magnitude of the resulting 2D map $\mathcal{F}_t[y]$ is extracted with integral normalized to 1 to obtain the final representation $\mathcal{C}[f_c, f_D]$, constituting a normalized Doppler-cadence magnitude map defined as:
    \begin{equation}
        \mathcal{C}[f_c, f_D] = \frac{|\mathcal{F}_t(y)]|}{\sum_{f_c=1}^{M}\sum_{f_D=1}^{N}|\mathcal{F}_t(y)|_{f_c,f_D}}
    \end{equation}
    
    An identical derivation applied to $\hat{y}[t, f_D]$ yields $\hat{\mathcal{C}}[f_c, f_D]$.
    
    Since $\mathcal{C}[f_c, f_D]$ and $\hat{\mathcal{C}}[f_c, f_D]$ are normalized 2D representations, the form of the $\mathcal{L}_{\mu\textrm{ D}}(y, \hat{y})$ loss term will be similar to \eqref{eq:pixel_loss}, since \gls{mse} is used to compare them:

    \begin{equation}
        \label{md_loss_final}
        \mathcal{L}_{\mu\textrm{ D}}(y, \hat{y}) = \frac{1}{N} \sum_{f_D=1}^{N} \underbrace{\sum_{f_c=1}^{M}\frac{(\mathcal{C}_{f_c,f_D}-\hat{\mathcal{C}}_{f_c,f_D})^2}{M}}_{\mathcal{S}[f_D]}
    \end{equation}  
    
    The new term in $\mathcal{L}_{\mu\textrm{ D}}$ in the objective function $J_\theta$ in \eqref{eq:j_updated} emphasizes features directly related to Micro-Doppler oriented tasks. The initial phase information for each oscillatory mode within a Doppler frequency bin is ignored by applying the magnitude operation. The periodicity of the Doppler components has to match the truth in order to minimise the micro-Doppler coherence loss. Furthermore, the loss is invariant to small shifts in time of individual micro-Doppler spectral components\footnote{However, significant shifts in time of individual components will be penalised by $\mathcal{L}_{\textrm{MSE}}$ due to the resulting pixel error.}.

\section{Evaluation}

    The utility of the micro-Doppler coherence loss is demonstrated using a dataset containing real radar signatures of various human activities. The dataset is publicly available\footnote{Available at \url{http://researchdata.gla.ac.uk/848/}}, containing 1,752 samples with ground truth~\cite{glasgowdataset}. The samples contain signatures from 6 different activities: 1) Walking, 2) Sitting down, 3) Standing up, 4) Object Pick Up, 5) Drinking, 6) Fall. The samples are divided into training, validation and test datasets with ratios of (0.5, 0.25, 0.25), respectively. Since the class imbalances in the dataset are not severe, no balancing countermeasures are applied.
    
    The general learning approach is similar to \cite{Seyfioglu2018a}, where classification training is preceded by an unsupervised stage enabling the evaluation in the improvement in both the domain translation as well as the classification contexts.
    
    The structure of the model is shown in Figure~\ref{fig:model_diagram}. The network utilises 128 by 128 spectrogram images with 2 channels to accommodate real and imaginary components. The spectrogram is computed with 128 bins in a 0.2 seconds Blackman window and 0.19 seconds overlap. The resulting spectrogram image is then uniformly sampled in time to obtain 128 spectra, yielding a 128 by 128 complex matrix. The model encoder translates this image to a latent code of size 128, as shown in Figure~\ref{fig:model_diagram}.  All convolutional layers in the network use a kernel of size 3 with a stride of 2. The latent code is then input to the decoder module; alternatively, the same code can be fed to the classifier module. The hybrid structure allows for convenient switching between the translation and the classification operation.

    The results rely on a comparison between a network where only \gls{mse} reconstruction loss $\mathcal{L}_{\mathrm{MSE}}$ is contained in the objective function $J_\theta$  and a network where the micro-Doppler loss term $\mathcal{L}_{\mu\mathrm{D}}$ is added to the objective with a weight $\beta$ of 4.

    \subsection{Unsupervised Stage}
    The influence of the proposed loss term can be demonstrated by investigating the loss curves of the trained convolutional auto-encoding model component. In the long term, the decay of the reconstruction loss can be expected to drive the micro-Doppler loss down also. However, the degree to which the two losses are coupled remains to be demonstrated. This stage also provides confirmation on whether using the additional loss term significantly changes the direction of gradients used for backpropagation.
    
    \begin{figure*}[t]
        \centering
        \begin{tabular}{c c c}
            \includegraphics[width=0.3\textwidth]{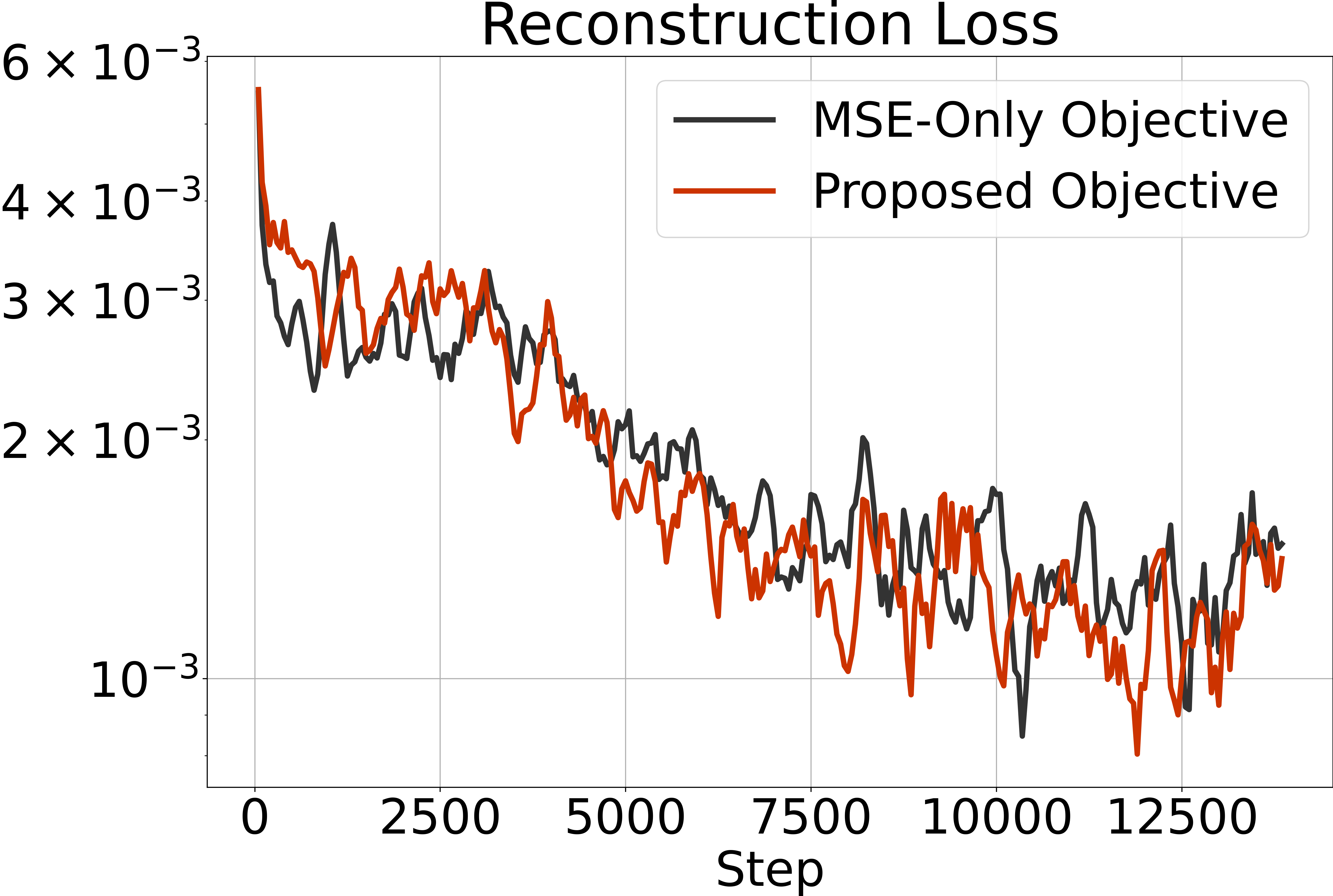} & \includegraphics[width=0.3\textwidth]{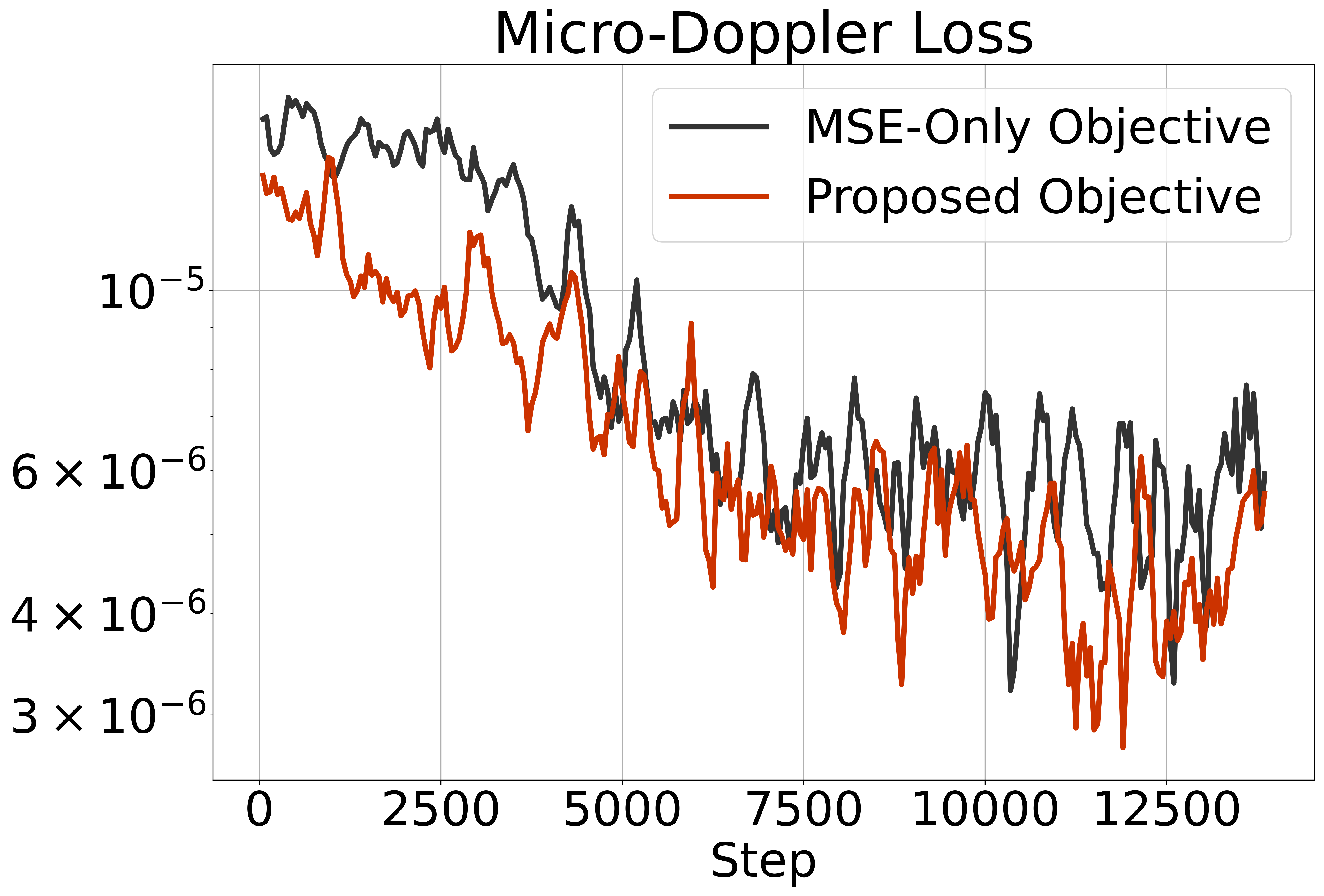} &
            \includegraphics[width=0.3\textwidth]{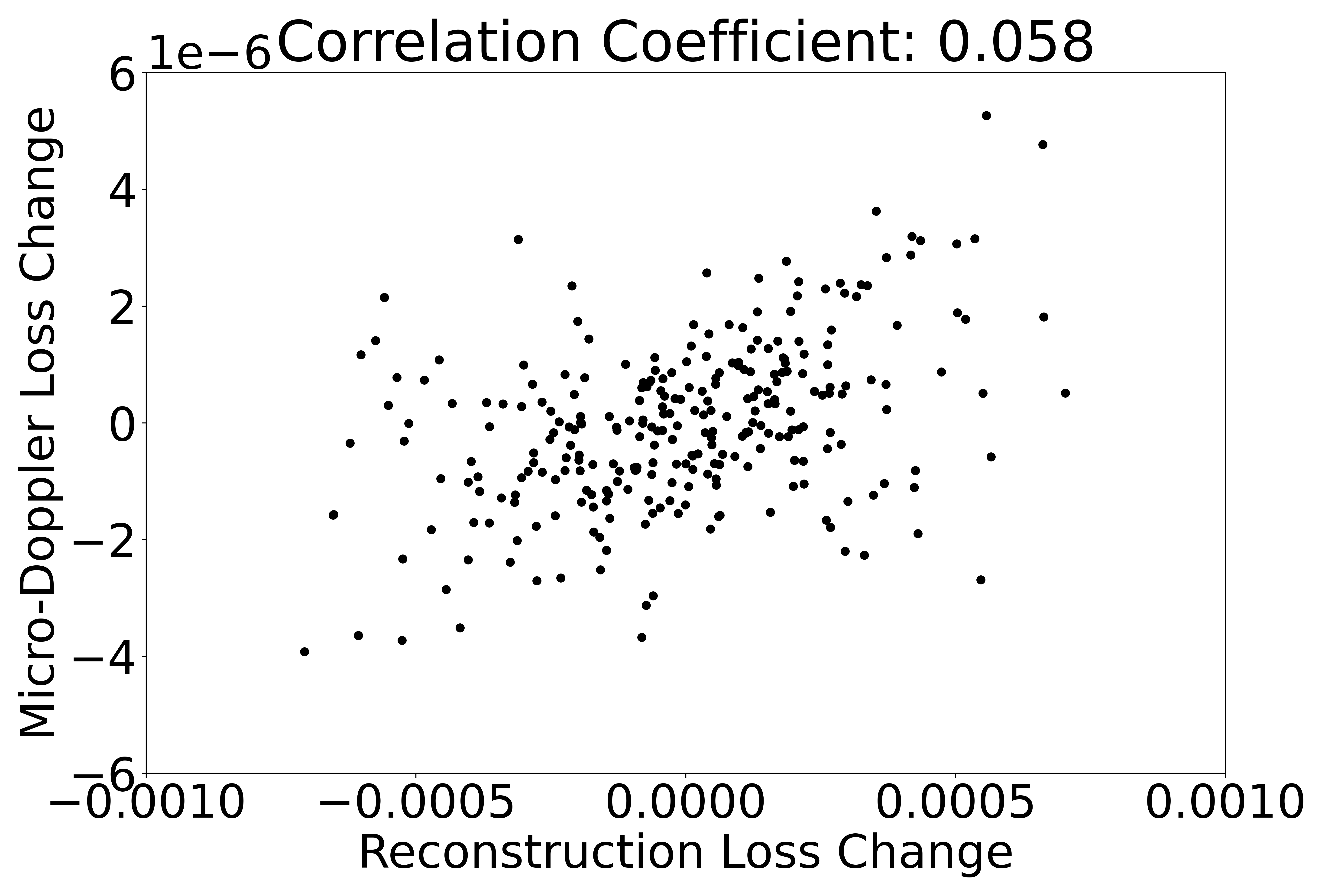}\\
            (a) & (b) & (c)
        \end{tabular}
        \caption{Comparison of autoencoding loss curves. (a) Reconstruction Loss (b) micro-Doppler Loss and (c) scatter plot of the Reconstruction Loss Change and micro-Doppler Loss Change}
        \label{fig:loss_curves}
    \end{figure*}
    
    \begin{figure}[!bh]
        \centering
        \begin{tabular}{c}
            \includegraphics[width=0.72\columnwidth]{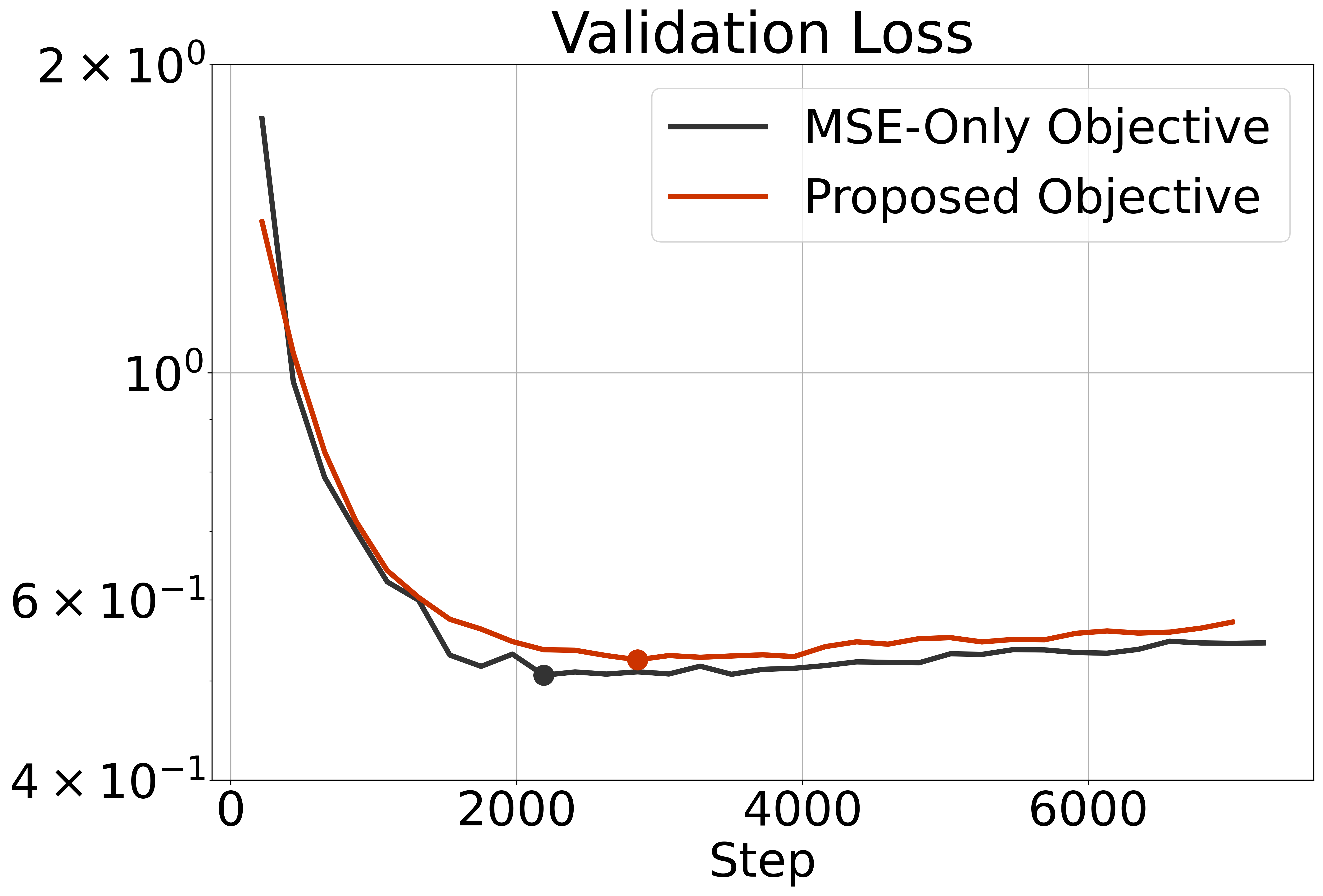}\\(a)\\
            \includegraphics[width=0.7\columnwidth]{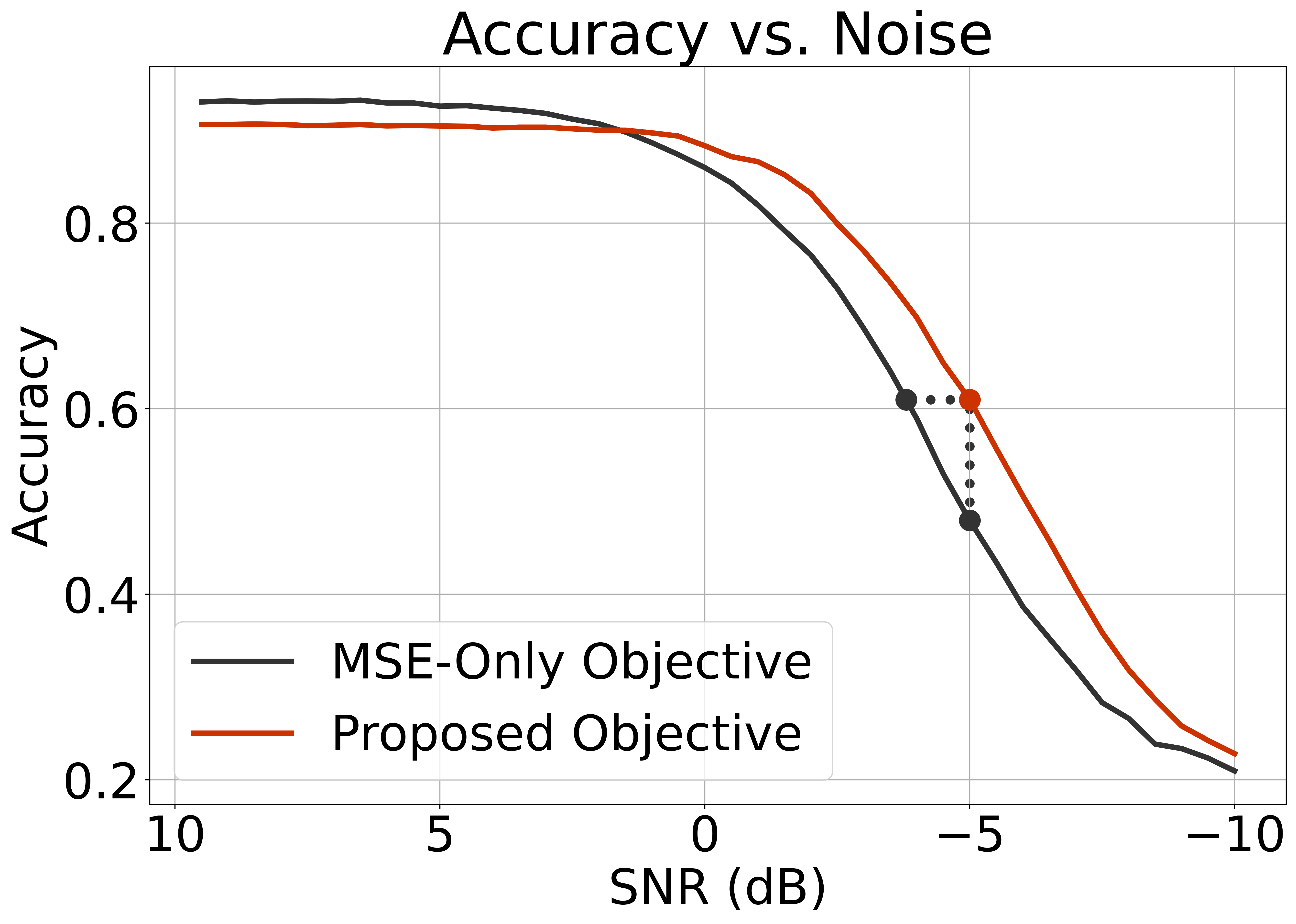} \\
             (b)
        \end{tabular}
        \caption{Comparison of the relationship between accuracy and \gls{snr} of tested samples}
        \label{fig:val_n_noise}
    \end{figure}
    
    Figure~\ref{fig:loss_curves} illustrates how both the reconstruction and micro-Doppler losses vary with each weight update. In the case of the \gls{mse}-only objective (black), both losses are reduced in the long term; however, the latter decays at a slower rate than in the case of the proposed objective function (red). Thus the reconstruction loss and the micro-Doppler loss gradients are only partially aligned, implying that each loss can be associated with a different set of learned features. Further confirmation can be obtained by investigating the correlation between the change in the reconstruction loss and the change in the micro-Doppler loss for the scenario where the objective function contains only the reconstruction loss (black). The scatter plot in Figure~\ref{fig:loss_curves}(c) illustrates that relationship. The correlation coefficient between the two variables is 0.058 suggesting no consistent relationship between them and confirming that the gradients propagated from the reconstruction loss generally point in a different direction than in the case of the proposed objective.
    
    \subsection{Classification}
    Two sets of model weights pre-trained in the unsupervised stage have been used subsequently to train a classifier head to discriminate between the three classes contained in the dataset (only cross-entropy loss is included in the objective function at this stage). Figure~\ref{fig:val_n_noise}(a) shows the validation curves for the two sets of pre-trained weights. Evident is that the weights obtained using the addition of micro-Doppler loss in the unsupervised stage (red) lead to a smoother validation loss decay than the standard \gls{mse}-only objective (black). The dots mark the lowest validation loss achieved in each case; 0.50647 for the standard approach, and 0.52433 for the proposed micro-Doppler coherence loss. The weights from these states have been extracted in order to compare the two approaches.
    
    Using the best-performing weights, the accuracy of the classifier has been tested against varying levels of additive white noise (\gls{snr} swept from 10 to -10 dB), as shown in Figure~\ref{fig:val_n_noise}(b). Results indicate that the application of the proposed micro-Doppler coherence loss yields a model more robust to noise compared to the conventional approach.
    
    \begin{figure}[b]
        \centering
        \begin{tabular}{c@{\hspace{0.01\columnwidth}}c}
            \includegraphics[width=0.48\columnwidth]{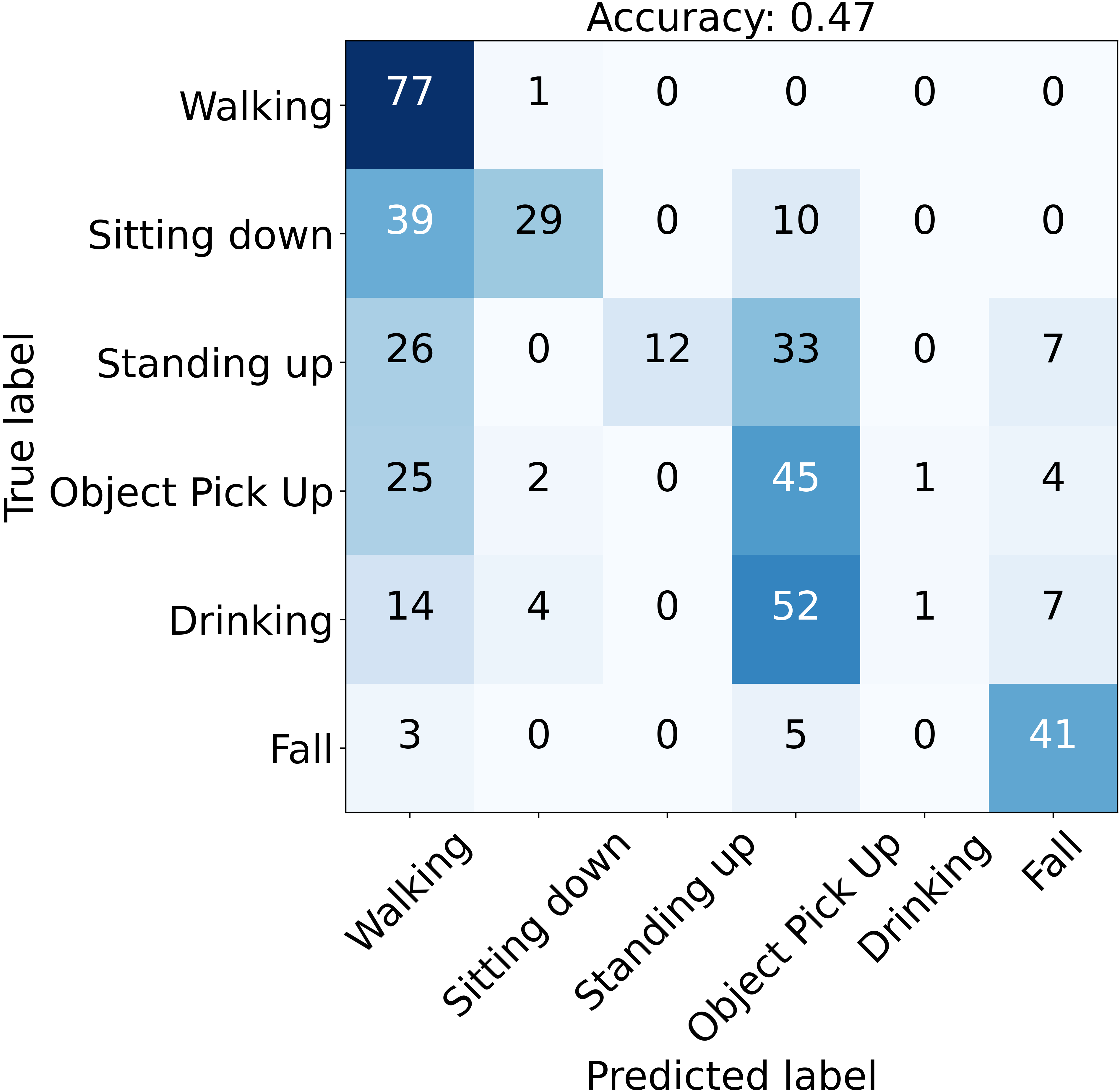} & \includegraphics[width=0.48\columnwidth]{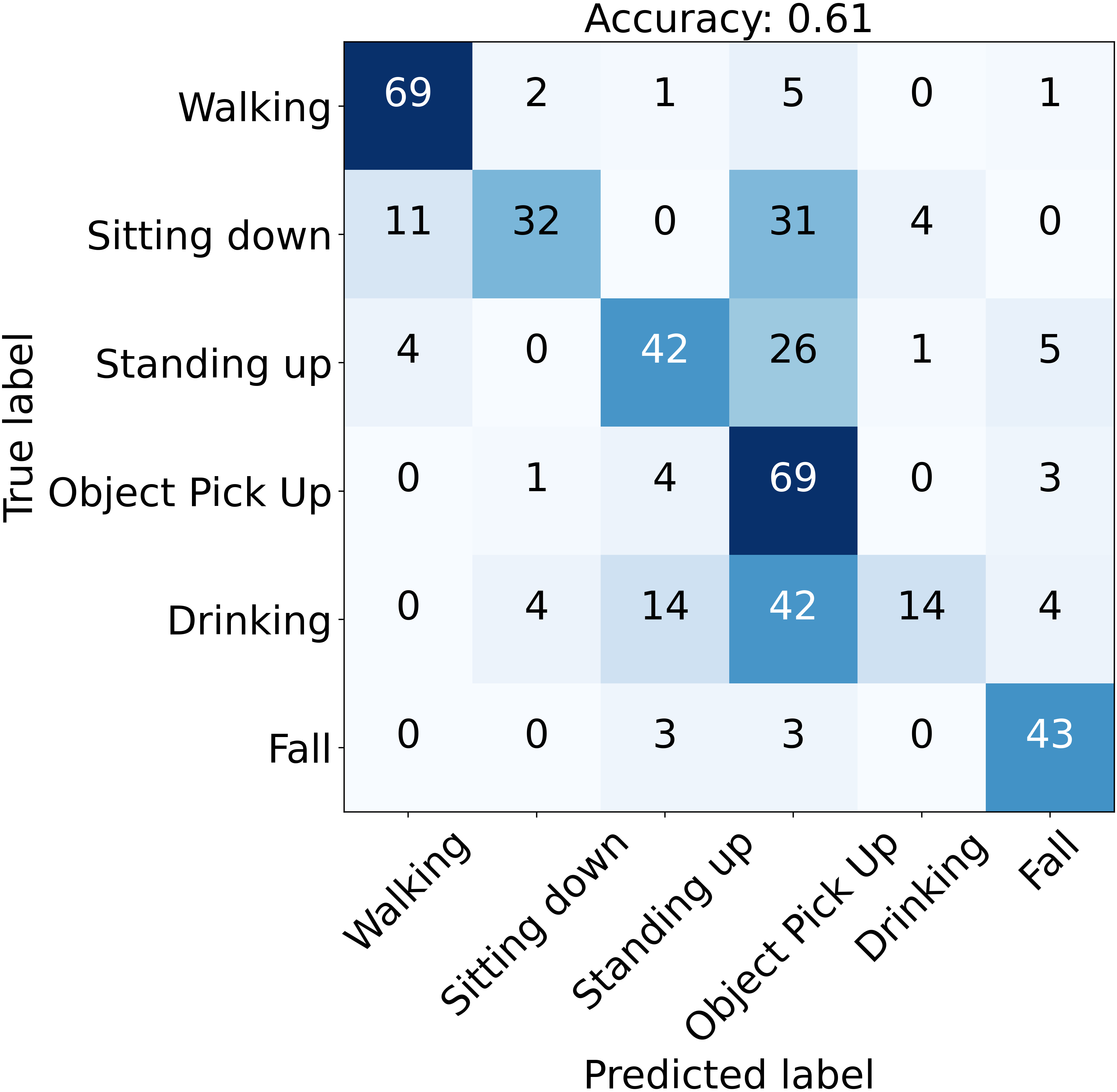} \\
            (a) & (b)
        \end{tabular}
        \caption{Comparison of obtained confusion matrix at -5 $\mathrm{SNR}_{\mathrm{dB}}$ for (a)~a~scenario with no micro-Doppler loss applied (b)~a~scenario with micro-Doppler loss backpropagation.}
        \label{fig:confusiom_matrices}
    \end{figure}
    
    The advantage gained in the context of the classification task is further observed in the confusion matrices for the level of noise where the difference of accuracy is most significant. The confusion matrices presented in Fig.~\ref{fig:confusiom_matrices} demonstrate the performance achieved by both networks with injected additive input noise of SNR equal to -5 dB. Further, the network output for each sample has been computed for 16 different noise samples in order to obtain a representative example. The proposed micro-Doppler coherence loss results in an increase in accuracy from 0.48 to 0.61 (marked by the two vertically aligned dots). Conversely, for a set accuracy level of 0.61, the proposed approach can accommodate an additional 1.2 dB of added noise with no drop in accuracy. The confusion matrix for the model trained using the micro-Doppler coherence loss is shown in Figure~\ref{fig:confusiom_matrices}(b). The number of correctly classified samples for almost all classes increase significantly compared to the standard approach in Figure~\ref{fig:confusiom_matrices}(a). The numbers of correct predictions for the Walking and the Fall class are lower for the proposed approach, however, the differences are minimal. Nevertheless, the total number of correct predictions is higher for the model trained with micro-Doppler coherence loss over a range of noise level values as illustrated in Figure\ref{fig:val_n_noise}(b).
    
\section{Conclusions}

    A novel coherence loss term has been proposed for training deep learning models operating on Doppler time-frequency representations. Inclusion of the loss term in the objective function provides more appropriate optimization gradients for micro-Doppler applications. Results indicate that this practice can be beneficial not only when the network output target is a time-frequency map but also in a classification framework. The new loss term utilised in the unsupervised pre-training stage leads to a classifier significantly more resilient to noise, making the model accuracy invariant to approximately 1.2 dB of additional noise, or 10 percentage points higher accuracy for the same level of noise is achieved.


\bibliographystyle{IEEEtran}
\bibliography{references}

\end{document}